\documentstyle[12pt]{article}
\begin{document}

\tolerance=5000

\def\pp{{\, \mid \hskip -1.5mm =}}
\def\cL{{\cal L}}
\def\be{\begin{equation}}
\def\ee{\end{equation}}
\def\bea{\begin{eqnarray}}
\def\eea{\end{eqnarray}}
\def\tr{{\rm tr}\, }
\def\nn{\nonumber \\}
\def\e{{\rm e}}
\def\D{{D \hskip -3mm /\,}}

\ 

\vskip -2cm

\ \hfill
\begin{minipage}{3.5cm}
NDA-FP-46 \\
April 1997 \\
\end{minipage}

\vfill

\begin{center}
{\Large\bf Quantum evolution of Schwarzschild-de Sitter (Nariai)
black holes }

\vfill

{\sc Shin'ichi NOJIRI}\footnote{\scriptsize 
e-mail: nojiri@cc.nda.ac.jp} and
{\sc Sergei D. ODINTSOV$^{\spadesuit}$}\footnote{\scriptsize 
e-mail: odintsov@tspi.tomsk.su}

\vfill

{\sl Department of Mathematics and Physics \\
National Defence Academy, 
Hashirimizu Yokosuka 239, JAPAN}

\ 

{\sl $\spadesuit$ 
Tomsk Pedagogical University, 634041 Tomsk, RUSSIA \\
}



\ 

\vfill

{\bf abstract}

\end{center}

We calculate the one-loop effective action for conformal 
matter (scalars, spinors and vectors) on spherically symmetric 
background. Such effective action (in large $N$ approximation 
and expansion on curvature) is used to study quantum aspects of
 Schwarzschild-de Sitter black holes (SdS BHs) in nearly 
degenerated limit (Nariai BH). We show that for all types of
above matter SdS BHs may evaporate or anti-evaporate in 
accordance with recent observation by Bousso and Hawking 
for minimal scalars. Some remarks about energy flow for 
SdS BHs in regime of evaporation or anti-evaporation are 
also done. Study of no boundary condition shows that this condition 
 supports  anti-evaporation for nucleated BHs (at least 
in frames of our approximation). That indicates to the possibility 
that some pair created cosmological BHs may not 
only evaporate but also anti-evaporate. Hence, cosmological primordial 
BHs may survive much longer than it is expected.

\ 


\newpage

\section{Introduction}

~~~~~~In the absence of consistent quantum gravity, the natural way to take 
into account quantum effects in the early Universe or in black holes (BHs) 
is to consider matter quantum field theory (say, some GUT) in curved 
background. 
The study of quantum GUTs in curved space-time (see \cite{BOS} for a review)
shows the existence of beautiful phenomenon -- asymptotic conformal 
invariance (see original works \cite{BO} or book \cite{BOS}, for a review).
According to it, there exists large class of asymptotically free GUTs 
which tend to conformally invariant free theory at high curvature or 
at high energies(i.e. in the vicinity of BH or in the early 
Universe). Hence, for above background one can describe GUT as the collection 
of free conformal fields. If one knows the effective action of such system 
one can apply it for investigation of quantum evolution of strongly 
gravitating objects.

In recent work \cite{BH} the quantum evolution of Schwarzschild-de Sitter 
(Nariai) BHs has 
been studied for Einstein gravity with $N$ minimal quantum scalars. Large $N$ 
and $s$-waves approximation has been used in such an investigation. 
The possibility of quantum anti-evaporation of such BHs 
(in addition to well-known 
evaporation process \cite{Ha}) has been discovered. In the paper \cite{NOa} 
another model (of quantum conformal scalars with Einstein gravity) has 
been considered in a better approach to effective action (large $N$ 
approximation, partial expansion on curvature and partial 
$s$-waves reduction). The possibility of Schwarzschild-de Sitter (SdS) BHs 
anti-evaporation has been confirmed as well in the model of ref.\cite{NOa}.

Having in mind above remarks on the representation of some GUT in
the vicinity of BHs as collection of free conformal fields, we continue 
to study quantum dynamics of SdS BHs. We start from Einstein gravity with
quantum conformal matter ($N$ scalars, $N_1$ vectors and $N_{1/2}$ fermions). 
Working in large $N$ approximation (where only matter quantum effects are 
dominant) we also use partial derivative expansion of EA
(without $s$-waves reduction). As a main qualitative result 
we find that extreme SdS (Nariai) BHs 
may indeed evaporate as well as anti-evaporate. We also try 
to answer the question: 
Can the no-boundary Hartle-Hawking condition be 
consistent with anti-evaporation? This question may be really 
important for estimation of primordial BHs creation(\cite{G} and refs. 
therein) (and their 
existence in the present Universe) as SdS BHs actually may appear 
through such a process.

\section{Effective Action for Conformal Matter}

~~~~~~We first derive the effective action for 
conformally invariant matter (for a general review of 
effective action in curved space, see\cite{BOS}).
Let us start from Einstein gravity with $N$ 
conformal scalars $\chi_i$, $N_1$ vectors $A_\mu$ and 
$N_{1/2}$ Dirac spinors $\psi_i$
\bea
\label{OI}
S&=&-{1 \over 16\pi G}\int d^4x \sqrt{-g_{(4)}}
\left\{R^{(4)} -2\Lambda\right\} 
+ \int d^4x \sqrt{-g_{(4)}}\left\{{1 \over 2}\sum_{i=1}^N 
\left(g_{(4)}^{\alpha\beta}\partial_\alpha\chi_i
\partial_\beta\chi_i \right.\right. \nn
&& \left.\left.+ {1 \over 6}R^{(4)}\chi_i^2 \right)
-{1 \over 4}\sum_{j=1}^{N_1}F_{j\,\mu\nu}F_j^{\mu\nu}
+\sum_{k=1}^{N_{1/2}}\bar\psi_k\D\psi_k \right\}\ .
\eea
The convenient choice for the spherically symmetric 
space-time is the following:
\be
\label{OIV}
ds^2=f(\phi)\left[
f^{-1}(\phi)g_{\mu\nu}dx^\mu dx^\nu + r_0^2 d\Omega\right]\ .
\ee
where $\mu,\nu=0,1$, $g_{\mu\nu}$ and $f(\phi)$ depend only 
from $x^0$, $x^1$ and $r_0^2$ is the constant.

Let us start the calculation of effective action due to 
conformal matter on the background (\ref{OIV}).  
In the calculation of effective action, we present 
effective action as : 
$\Gamma=\Gamma_{ind}+ \Gamma[1, g^{(4)}_{\mu\nu}]$
where $\Gamma_{ind}=\Gamma[f, g^{(4)}_{\mu\nu}]
- \Gamma[1, g^{(4)}_{\mu\nu}]$ is conformal anomaly 
induced action which is quite well-known \cite{R}, 
$g^{(4)}_{\mu\nu}$ is metric (\ref{OIV}) without 
multiplier in front of it, i.e., $g^{(4)}_{\mu\nu}$ 
corresponds to 
\be
\label{OVI}
ds^2=\left[
\tilde g_{\mu\nu}dx^\mu dx^\nu + r_0^2 d\Omega\right]\ ,\ \ \ 
\tilde g_{\mu\nu}\equiv f^{-1}(\phi)g_{\mu\nu} \ .
\ee

The conformal anomaly for above matter is well-known 
\be
\label{OVII}
T=b\left(F+{2 \over 3}\Box R\right) + b' G + b''\Box R
\ee
where $b={(N +6N_{1/2}+12N_1)\over 120(4\pi)^2}$, 
$b'=-{(N+11N_{1/2}+62N_1) \over 360(4\pi)^2}$, $b''=0$ but 
in principle, $b''$  may be changed by the finite 
renormalization of local counterterm in gravitational 
effective action, $F$
is the square of Weyl tensor, $G$ is 
Gauss-Bonnet invariant.

Conformal anomaly induced effective action $\Gamma_{ind}$ 
may be written as follows \cite{R}:
\bea
\label{OVIII}
W&=&b\int d^4x \sqrt{-g} F\sigma 
+b'\int d^4x \sqrt{-g} \Bigl\{\sigma\left[
2\Box^2 + 4 R^{\mu\nu}\nabla_\mu\nabla_\nu \right. \nn
&& \left. - {4 \over 3}R\Box + {2 \over 3}(\nabla^\mu R)\nabla_\mu 
\right]\sigma + \left(G-{2 \over 3}\Box R\right)\sigma \Bigr\} \nn
&& -{1 \over 12}\left(b'' + {2 \over 3}(b + b')\right)
\int d^4x \sqrt{-g}\left[R - 6 \Box \sigma 
- 6(\nabla \sigma)(\nabla \sigma) \right]^2 
\eea
where $\sigma={1 \over 2}\ln f(\phi)$, and 
$\sigma$-independent terms are dropped. All 4-dimensional 
quantities (curvatures, covariant derivatives) in 
Eq.(\ref{OVIII}) should be calculated on the metric 
(\ref{OVI}). (We did not write subscript $(4)$ for them.)
Note that after calculation of (\ref{OVIII}) on the metric 
(\ref{OVI}), we will get effectively two-dimensional 
gravitational theory.

In the next step, we are going to calculate 
$\Gamma[1, g^{(4)}_{\mu\nu}]$. 
This term corresponds to the 
conformally invariant part of effective action. In this 
calculation, we may apply Schwinger-De Witt (SDW) 
type expansion of effective action with zeta-regularization\cite{E} 
(or other ultraviolet regularization). 
This expansion represents the expansion on powers of curvature 
invariants. Note that we 
add such EA to Einstein gravity action. Hence, two first terms 
of the SDW expansion (cosmological and linear curvature terms)
may be dropped as they only lead to finite renormalization of 
Hilbert-Einstein action (redefinition of cosmological and 
gravitational coupling constants). 
Then the leading (curvature quadratic term of this expansion) 
may be read off (see \cite{BOS}) as follows:
\be
\label{SDW}
\Gamma[1, g_{\mu\nu}^{(4)}]=\int d^4x 
\sqrt{-g}\left\{\left[{b}F+  {b'}G
+ { 2b\over 3}\Box R\right]\ln { R \over \mu^2}\right\}
+{\cal O}(R^3)
\ee
where $\mu$ is mass-dimensional constant parameter, all 
the quantities are calculated on the background (\ref{OVI}). 
The condition of application of above expansion is 
$|R|<R^2$ (curvature is nearly constant), in this case 
we may limit by only few first terms.

\section{Quantum Dynamics on Spherical Background}

~~~~~~We now consider to solve the equations of motion obtained from 
the above effective Lagrangians $S+\Gamma$. 
In the following, we use $\tilde g_{\mu\nu}$ and $\sigma$ 
as a set of independent variables and we write 
$\tilde g_{\mu\nu}$ as $g_{\mu\nu}$ if there is no 
confusion.

$\Gamma_{ind}$ ($W$ in Eq.(\ref{OVIII})) is rewritten after the reduction 
to 2 dimensions as follows:
\bea
\label{Gindrd}
{\Gamma_{ind} \over 4\pi}&=&
{b r_0^2 \over 3}\int d^2x\sqrt{-g}\left(
\left(R^{(2)} + R_\Omega\right)^2 + {2 \over 3}R_\Omega R^{(2)} 
+ {1 \over 3}R_{\Omega}^2 \right) \sigma \nn
&& + b' r_0^2 \int d^2x\sqrt{-g}\left\{\sigma \left(
2\Box^2 + 4R^{(2)\mu\nu}\nabla_\mu \nabla_\nu 
-{4 \over 3}(R^{(2)} + R_\Omega)\Box \right.\right. \nn
&& \left. 
+ {2 \over 3}(\nabla^\mu R^{(2)})\nabla_\mu\right) 
\sigma \left. + \left( 2R_\Omega R^{(2)} - {2 \over 3}\Box R^{(2)}
\right)\sigma \right\} \nn
&& -{1 \over 12}\left\{ b'' + {2 \over 3}(b+b')\right\}r_0^2
\int d^2x\sqrt{-g} \nn
&& \times \left\{ \left( R^{(2)} + R_\Omega - 6 \Box\sigma 
- 6\nabla^\mu\sigma\nabla_\mu\sigma \right)^2 
- \left( R^{(2)} + R_\Omega \right)^2 \right\} \eea
Here $R_\Omega={2 \over r_0^2}$ is scalar curvature of $S^2$ with 
the unit radius. The suffix ``(2)'' expresses the quantity in 2 
dimensions but we abbreviate it if there is no any confusion.
We also note that in two dimensions the Riemann tensor 
$R_{\mu\nu\sigma\rho}$ and $R_{\mu\nu}$ are expressed via the scalar 
curvature $R$ and the metric tensor $g_{\mu\nu}$ as : 
$R_{\mu\nu\sigma\rho}={1 \over 2}\left(g_{\mu\sigma}g_{\nu\rho}
- g_{\mu\rho}g_{\nu\sigma}\right)R$ and 
$R_{\mu\nu}={1 \over 2}g_{\mu\nu}R$.

Let us derive the equations of motion 
with account of quantum corrections from above effective action.
In the following, we work in  the conformal gauge :
$g_{\pm\mp}=-{1 \over 2}\e^{2\rho}\ ,\ \ \ g_{\pm\pm}=0$
after considering the variation of the effective action 
$\Gamma+S$ with respect to $g_{\mu\nu}$ and $\sigma$.
Note that the tensor $g_{\mu\nu}$ under consideration is 
the product of the 
original metric tensor and the $\sigma$-function 
$\e^{-2\sigma}$, the equations given by the variations of 
$g_{\mu\nu}$ are the combinations of the equations given by the 
variation of the original metric and $\sigma$-equation.

It often happens that we can drop the terms linear in $\sigma$ 
in (\ref{Gindrd}). In particular, one can redefine the corresponding 
source term as it is in the case of IR sector of 4D QG \cite{AMO}. 
In the following, we only consider this case. 
Then the variation of $S+\Gamma_{ind}+\Gamma[1, g^{(4)}_{\mu\nu}]$
with respect to $g^{\pm\pm}$ is given by
\bea
\label{cons1}
0&=&{1 \over 4\pi}{\delta (S+\Gamma_{ind}+\Gamma[1, g^{(4)}_{\mu\nu}]) 
\over \delta g^{\pm\pm}} \nn
&=&-{ r_0^2 \over 16\pi G}\e^{2\rho+2\sigma}\left[(\partial_\pm\sigma)^2 
-\partial_\pm^2\sigma+2\partial_\pm\sigma\partial_\pm\rho\right] 
+ b' r_0^2\left[ 8\e^{2\rho}\partial_\pm \sigma \partial_\pm 
\left(\e^{-2\rho}\partial_+\partial_- \sigma \right) \right. \nn
&& \left. -8\sigma\partial_\pm^2\sigma \partial_+\partial_- \rho 
+ {2 \over 3}\e^{2\rho}\partial_\pm\sigma \partial_\pm\left\{R_4
\sigma \right\} + {8 \over 3}\e^{2\rho}\sigma 
\partial_\pm R_4\partial_\pm \sigma  \right] \nn
&& -\left\{b''+{2 \over 3}(b+b')\right\} r_0^2\left[
4e^{2\rho}\partial_\pm\sigma \partial_\pm R_4 
- 4 (\partial_\pm \sigma)^2\partial_+\partial_-\rho \right. \nn
&& \left. + 12 e^{2\rho}\partial_\pm\sigma \partial_\pm
\{\e^{-2\rho}(\partial_+\partial_-\sigma 
+ \partial_+ \sigma \partial_-\sigma)\} -12 (\partial_+\partial_-\sigma 
+ \partial_+ \sigma \partial_-\sigma)(\partial_\pm\sigma)^2 \right] \nn
&& + \left\{\left(- \partial_\pm^2\rho - 2(\partial_\pm\rho)^2
\right) -{1 \over 4} \partial_\pm^2 + {3 \over 2}\partial_\pm^2\rho 
+ {3 \over 2} \partial_\pm\rho \partial_\pm \right\} \nn
&& \times \left[ {16 \over 3}b' r_0^2(-\sigma\partial_+\partial_-\sigma 
+ \partial_+\sigma\partial_-\sigma) 
-\left\{b''+{2 \over 3}(b+b')\right\}  r_0^2
(\partial_+\partial_-\sigma + \partial_+\sigma\partial_-\sigma) \right]\nn
&& +r_0^2 \left[-{1 \over 3}b \e^{2\rho}\partial_\pm\left\{\ln \left( 
{R_4 \over \mu^2 } \right)\right\} \partial_\pm R_4 \right. \nn
&& + 4\left\{\left(- \partial_\pm^2\rho - 2(\partial_\pm\rho)^2
\right) -{1 \over 4} \partial_\pm^2 + {3 \over 2}\partial_\pm^2\rho 
+ {3 \over 2} \partial_\pm\rho \partial_\pm \right\} \nn
&& \times \left\{{8 \over 3}b\partial_+\partial_-\rho \ln \left( 
{R_4 \over \mu^2 } \right) \right. 
+ b\partial_+\partial_-\left\{\ln \left({R_4 \over\mu^2}\right)\right\} \\
&& \left. + \left\{b\left({32 \over 3}\e^{-2\rho}(\partial_+\partial_-\rho)^2
+ {2\e^{2\rho} \over 3 r_0^2} - {4 \over 3}\partial_+\partial_-R_4
+{16 \over r_0^2}\left({b \over 3}+b'\right)\right\} 
{1 \over R_4}\right\} \right] \nonumber
\eea
Here $R_4\equiv 8\e^{-2\rho}\partial_+\partial_-\rho + {2 \over r_0^2}$.
Usually the equation given by $g^{++}$ or $g^{--}$ can be regarded as 
the constraint equation with respect to the initial or boundary conditions.
The equations obtained here are, however, combinations of the constraint 
and $\sigma$-equation of the motion since the tensor $g_{\mu\nu}$ 
under consideration is the product of the original metric tensor and 
the $\sigma$-function $\e^{-2\sigma}$.

The variations with respect to $\rho$ are given by 
\bea
\label{rhov}
0&=&{1 \over 4\pi}
{\delta (S+\Gamma_{ind}+\Gamma[1, g^{(4)}_{\mu\nu}]) \over \delta \rho} \nn
&=& -{ r_0^2 \over 16\pi G} \left[4\partial_+\partial_-\e^{2\sigma}
-2\e^{2\rho+4\sigma}\Lambda + {4 \over r_0^2}\e^{2\rho+2\sigma}\right] 
+b' r_0^2 \left\{ -32(\partial_+\partial_-\sigma)^2\e^{-2\rho} \right.\nn
&&-{128 \over 3}\e^{-2\rho} \partial_+\partial_-\rho 
(\sigma\partial_+\partial_-\sigma) 
+ {64 \over 3}\partial_+\partial_-\left(\e^{-2\rho}
\sigma\partial_+\partial_-\sigma\right) \nn
&& - {16 \over 3}\left\{-2\partial_+\sigma\partial_-\sigma
\e^{-2\rho}\partial_+\partial_-\rho 
+ \partial_+\partial_-\left(\partial_+\sigma\partial_-\sigma
\e^{-2\rho}\right)\right\} \nn
&& -\left\{b''+ {2 \over 3}(b+b')\right\} r_0^2\left[
16 \partial_+\partial_- \left\{ \e^{-2\rho} (\partial_+\partial_-\sigma
+\partial_+\sigma\partial_-\sigma)\right\} \right. \nn
&& \left. -48\e^{-2\rho}(\partial_+\partial_-\sigma
+\partial_+\sigma\partial_-\sigma)^2\right] \nn
&& +r_0^2\left[ -{64 \over 3}b\e^{-2\rho}\left(\partial_+\partial_-
\rho \right)^2 \ln \left( {R_4 \over \mu^2 } \right) \right. 
+{64 \over 3}b\partial_+\partial_-\left\{\e^{-2\rho}\partial_+\partial_-
\rho \ln \left( {R_4 \over \mu^2 } \right)\right\} \nn
&& + {4b\e^{2\rho} \over 3 r_0^2} \ln \left({R_4 \over \mu^2 } \right) 
+{16 \over r_0^2}\left({b \over 3}+b'\right)\partial_+\partial_- 
\ln \left({R_4 \over \mu^2 } \right) \\
&& +{64 \over 3}b\e^{-2\rho}\partial_+\partial_-\rho \partial_+\partial_- 
\left\{\ln \left( {R_4 \over \mu^2 } \right)\right\} 
-{64 \over 3}b\partial_+\partial_-\left\{\e^{-2\rho} \partial_+\partial_- 
\left\{\ln \left( {R_4 \over \mu^2 } \right)\right\}\right\} \nn
&& - {16\e^{-2\rho}\partial_+\partial_-\rho \over R_4 } 
\left\{{32 \over 3}b\e^{-2\rho}\left(\partial_+\partial_-\rho\right)^2
+ {2b\e^{2\rho} \over 3 r_0^2} 
- {4 \over 3}b\partial_+ \partial_-R_4
+ {16 \over r_0^2}\left({b \over 3}+b'\right)\partial_+\partial_-
\rho \right\} \nn
&& - \partial_+\partial_-\left\{{8\e^{-2\rho}\over R_4 }
\left\{{32 \over 3}b\e^{-2\rho}
(\partial_+\partial_-\rho)^2+ {2b\e^{2\rho} \over 3r_0^2} 
\left. - {4 \over 3}b\partial_+\partial_-R_4+ {16 \over r_0^2}
\left({b \over 3}+b'\right)
\partial_+\partial_-\rho \right\}\right\}\right] \nonumber
\eea
The variations with respect to $\sigma$  may be found as
\bea
\label{sigmav}
0&=&{1 \over 4\pi}
{\delta (S+\Gamma_{ind}+\Gamma[1, g^{(4)}_{\mu\nu}]) \over \delta \sigma} \nn
&=& -{ r_0^2 \over 16\pi G}
\Biggl[8\e^{2\sigma}\left\{3\partial_+\partial_- 
\sigma
+3\partial_+\sigma\partial_-\sigma + \partial_+\partial_-\rho \right\} 
-4\e^{2\rho+4\sigma}\Lambda + {4 \over r_0^2}\e^{2\rho+2\sigma} \Biggr] \nn
&& + b' r_0^2\left[ 32 \partial_+\partial_-(
\e^{-2\rho}\partial_+\partial_-\sigma) 
+{64 \over 3}\left(\e^{-2\rho}\partial_+\partial_-\rho
\partial_+\partial_-\sigma + \partial_+\partial_-(\sigma\e^{-2\rho}
\partial_+\partial_-\rho)\right) \right. \nn
&& \left. +{32 \over 3 r_0^2}\partial_+\partial_-\sigma 
+ {2 \over 3}\left\{\partial_+R_4\partial_-\sigma 
+\partial_-R_4\partial_+\sigma\right\} \right]\nn
&& -\left\{b''+ {2 \over 3}(b+b')\right\} r_0^2\left[
2\partial_+\partial_- R_4 
-2\left\{\partial_+ R_4\partial_-\sigma +\partial_-R_4
\partial_+\sigma\right\} \right. \nn
&& +48\partial_+\partial_-\left\{ \e^{-2\rho}
(\partial_+\partial_-\sigma +\partial_+\sigma\partial_-\sigma)\right\} 
-48\partial_+\left\{ \e^{-2\rho}\partial_-\sigma(\partial_+\partial_-
\sigma +\partial_+\sigma\partial_-\sigma)\right\} \nn
&& \left.-48\partial_-\left\{\e^{-2\rho}\partial_+\sigma(\partial_+
\partial_-\sigma+\partial_+\sigma\partial_-\sigma)\right\} \right] 
\eea
Note that now the real 4 dimensional metric is given by
: $ds^2=-\e^{2\sigma + 2\rho}dx^+dx^- + r_0^2\e^{2\sigma}d\Omega$.
Above equations give the complete system of quantum corrected 
equations of motion for the system under discussion.

\section{ Evolution of Schwarzschild-de Sitter Black Holes due to
Quantum Conformal Matter Back-Reaction}

~~~~~~We now consider the Schwarzschild-de Sitter family of black holes and 
its 
nearly degenerated case, so-called Nariai solution \cite{N}. We here follow 
the work by Bousso-Hawking \cite{BH}. The Schwarzschild type black 
hole solution in de Sitter space has two horizons, one is the usual 
event horizon and another one is the cosmological horizon, which is proper 
one  in de Sitter space. The Nariai solution is given by a limit of 
the Schwarzschild-de Sitter black hole where two horizons coincide 
with each other. In the limit, the two horizons have the same temperature 
since the temperature is proportional to the inverse root of the horizon
area. Therefore two horizons are in the thermal equilibrium in the limit. 
We are now interesting in the instability of the Nariai limit. Near the 
limit the temperature of the event horizon is higher than that of the 
cosmological one since the area of the event horizon is smaller than 
that of the cosmological horizon. This implies that there would be a 
thermal flow from the event horizon to the cosmological one. This means 
the system would become instable and the black hole would evaporate. We 
also have to note that above cosmological black holes may naturally 
appear through quantum pair creation \cite{GP,G} which may occur in 
the inflationary universe \cite{BH2}.

In the Nariai limit, the space-time has the topology of $S^1\times S^2$ and 
the metric is given by
\be
\label{Nsol}
ds^2={1 \over \Lambda}\left( \sin^2 \chi d\psi^2 - d\chi^2 - d\Omega
\right)\ .
\ee
Here coordinate $\chi$ has a period $\pi$. If we change the coordinates 
variables by
\be
\label{cv}
r=\ln\,\tan{\chi \over 2}\ ,\ \ \ t={\psi \over 4}\ ,
\ee
we obtain 
\be
\label{Nsol2}
ds^2= {1 \over \Lambda\cosh^2 r}\left(-dt^2 + dr^2\right) 
+ {1 \over \Lambda}d\Omega\ .
\ee
This form corresponds to the conformal gauge in two dimensions.
Note that the transformation (\ref{cv}) has one to one correspondence 
between $(\psi,\chi)$ and $(t,r)$ if we restrict $\chi$ by 
$0\leq \chi <\pi$ ($r$ runs from $-\infty$ to $+\infty$). 

Now we solve the equations of motion. Since the Nariai solution 
is characterized by the constant $\phi$(or $\sigma$),
we now assume that $\sigma$ is a constant even when including quantum 
correction
\be
\label{consigma}
\sigma=\sigma_0\ \ \mbox{(constant)}\ .
\ee
We also first consider static solutions and replace $\partial_{\pm}$ 
by $\pm{1 \over 2}\partial_r$. Then we find that the total constraint 
equation obtained by (\ref{cons1}) is trivially satisfied. 

 Assuming a solution is given by a constant 2d scalar curvature
\be
\label{conscur}
R=-2\e^{-2\rho}\partial_r^2\rho=R_0\ \ 
(\mbox{constant})\ ,
\ee
the equations of motions given by  (\ref{rhov}) (variation over  
 $\rho$) and (\ref{sigmav}) (variation over $\sigma$) become 
the following two algebraic equations
\bea
\label{aleq2}
0&=& -{r_0^2 \over 8\pi G}\left(-\Lambda\e^{4\sigma_0} 
+{2 \over r_0^2}\e^{2\sigma_0}\right) 
+r_0^2\left\{b\left(-{R_0^2 \over 3}+{4 \over 3r_0^4}\right)\ln\left(
{R_0 + {2 \over r_0^2} \over \mu^2}\right)\right. \nn
&& \left.-\left\{b\left({R_0^2 \over 3}
+{4 \over 3 r_0^4}\right)+ {8 \over r_0^2}\left({b \over 3}
+b'\right)R_0\right\}{R_0 \over R_0  + {2 \over r_0^2} }\right\} \\
\label{aleq1}
0&=&R_0 - 4\Lambda \e^{2\sigma_0} + {4 \over r_0^2} \ .
\eea
The above equations can be solved with respect to $\sigma_0$ and $R_0$ 
in general, although it is difficult to get the explicit expression. 

Equation (\ref{conscur}) can be integrated to be
\be
\label{rho}
\e^{2\rho}=\e^{2\rho_0}
\equiv{2C \over R_0}\cdot{1 \over \cosh^2 \left(r\sqrt{C}\right)} .
\ee
Here $C$ is a constant of integration.

We now consider the perturbation around the Nariai 
type solution (\ref{consigma}) and (\ref{rho})
\be
\label{pert}
\rho=\rho_0 + \epsilon R(t,r)\ ,\ \ \ 
\sigma=\sigma_0 + \epsilon S(t,r) \ .
\ee
Here $\epsilon$ is a infinitesimally small parameter.
Then we obtain the following linearized $\rho$ and $\sigma$ equations
\bea
\label{pertrho}
0&=&4\pi b' r_0^2\left\{-{16 \over 3}R_0\sigma_0\Delta S 
+{32 \over 3}{R_0\sigma_0 \over C}\Delta \left(\Delta S\right) \right\}
-4\pi(b+b') r_0^2{16 \over 3}{R_0 \over C}\Delta 
\left( \Delta S\right)  \nn
&& -{r_0^2 \over 16\pi G}\e^{2\sigma_0}\left[ 8\Delta S 
- {8\Lambda C \over R_0} \e^{2\sigma_0}(R+2S) + {16 C \over r_0^2 R_0} 
(R+S) \right] \nn
&& +4\pi r_0^2 \left[{2C \over R_0}
\left\{-{2bR_0 \over 3} \ln\left({R_4 \over \mu^2}\right)
+{b \over R_4}\left(- R_0^2 + {4 \over 3 r_0^4} \right) 
+ b\left({R_0^2 \over 3} + {4 \over 3 r_0^4} \right)
{2 \over r_0^2R_4^2} \right.\right. \nn
&& \left.-{8 \over r_0^2}\left({b \over 3}+b'\right)\left\{{4R_0 \over 
R_4} - {2R_0^2 \over R_0^2}\right\}\right\} 
\left(-2R_0R +  {4 R_0 \over C} \Delta R \right)  \nn
&& +\left\{ {8b \over 3}\ln\left({R_4 \over \mu^2}\right)
+{32b \over 3}{R_0 \over R_4 } 
+{4b \over R_4^2 }\left({R_0^2 \over 3} + {4 \over 3r_0^4} \right)\right\} 
\Delta \left(-2R_0R +  {4 R_0 \over C} \Delta R \right) \nn
&& - {32b \over 3}{R_0 \over CR_4}
\Delta \left(\Delta \left(-2R_0R +  {4 R_0 \over C} \Delta R 
\right) \right) \\
&& \left. + \left\{b\left(-{4R_0 \over 3}+{16 \over 3 r_0^4 R_0}\right)
\ln\left({R_4 \over \mu^2}\right)
-{4b \over 3}{R_0^2 \over R_4 } 
-{8b \over 3r_0^2}{2R_0 \over R_4}+{32 \over
r_0^2}\left({b \over 3} + b'\right){2R_0^2\over R_4}\right\}
CR_0 \right] \nn
\label{pertsigma}
0&=&4\pi b'r_0^2\left\{{16 \over 3}R_0 \Delta S +16{R_0 \over C}
\Delta \left(\Delta S\right) 
+ {32 \over 3 r_0^2} \Delta S + {8 \over 3}\sigma_0 
\Delta \left(-2R_0R +  {4 R_0 \over C} \Delta R \right) \right\} \nn
&& - 4\pi\times 16(b+b') r_0^2{R_0 \over C} \Delta \left(
\Delta S\right) 
- { r_0^2 \over 16\pi G}\e^{2\sigma_0}\left\{8\Delta 
S + \Delta R + {C \over 2} S \right. \nn
&& \left. -{16\Lambda C \over R_0} \e^{2\sigma_0} 
(R+2S)+ {16 C \over r_0^2 R_0} (R+S)\right\}
\eea
Here
\be
\label{Delta}
\Delta=\cosh^2\left(r\sqrt{C}\right)\partial_+\partial_-
\ee
and $R_4$ becomes constant $R_4=R_0 + {2 \over r_0^2}$.
The equations (\ref{pertrho}) and (\ref{pertsigma}) can be 
solved by assuming that $R$ and $S$ are given by the eigenfunctions of 
$\Delta$:
\be
\label{eigenRS}
R(t,r) =P f_A(t,r)\ ,\ \ \ S(t,r)= Q f_A(t,r)\ ,\ \ \  
\Delta f_A(t,r) = A f_A(t,r)\ .
\ee 
Note that $\Delta$ can be regarded as the Laplacian on the two dimensional
hyperboloid and the explicit expression for the eigenfunctions is given
later. Using (\ref{eigenRS}), we can rewrite (\ref{pertrho}) and 
(\ref{pertsigma}) (using (\ref{aleq1})) as follows:
\bea
\label{perte}
0&=& \left\{ {64\pi b' \sigma_0 R_0 \over 3}\left( - A+ 2 {A^2 \over C}
\right) - {64\pi (b+b') R_0 \over 3}{A^2 \over C} 
- {1 \over 4\pi G}\e^{2\sigma_0}\left(2A - C \right) \right\}Q \nn
&& + \left\{4\pi\left[ {2C \over R_0}\left\{-{2R_0 \over 3} 
\ln\left({R_4 \over \mu^2}\right)
+{b \over R_4}\left(-R_0^2 + {4 \over 3r_0^4} \right){ 1 \over R_4 }
+ {b \over R_4^2 } \left({R_0^2 \over 3} + {4 \over 3 r_0^4} \right)
{2 \over r_0^2} \right.\right.\right. \nn
&& \left. -{2 \over r_0^2}
\left({b \over 3}+b'\right)\left\{{4R_0 \over R_4} - 
{2R_0^2 \over R_4^2}\right\}\right\} 
\times \left(-2R_0 +  {4 R_0 \over C} A \right)  \nn
&& +\left\{ {8b \over 3}\ln\left({R_4 \over \mu^2}\right)
+{32b \over 3}{R_0 \over R_4 } 
+ \left({R_0^2 \over 360} + {1 \over 90 r_0^4} \right)
{ 4 \over R_4^2 }+{16 \over r_0^2R_4}\left(
{b \over 3}+b'\right)\right\} \nn 
&& \times \left(-2R_0A +  {4 R_0 \over C} A^2 \right) 
- {32bR_0 \over 3CR_4}\left(-2R_0
A^2 +  {4 R_0 \over C} A^3 \right) \nn
&& + \left\{b\left(-{4R_0 \over 3}+{16 \over 3 r_0^4 R_0}\right)\ln\left(
{R_4 \over \mu^2}\right)
-{4bR_0^2 \over 3R4} \right. \nn
&& \left.\left.-{8b \over 3r_0^2}{2 \over R_4}
-{8 \over r_0^2}\left({b \over 3}+b'\right){2R_0^2 \over R_4}
\right\}C \right]  
\left. - {C \over 8\pi G}\e^{2\sigma_0}\left( -1 + 
{4 \over r_0^2 R_0}\right)\right\}P \nn
&\equiv& M_Q(A) Q + M_P(A) P\ ,\nn
0&=&\left\{ 64\pi b' R_0 \left( {1 \over 3}A + {1 \over C}A^2 \right) 
+ {128
\pi \over 3 r_0^2}b'A - 64\pi (b+b'){R_0 \over C}A^2 \right. \nn
&& \left. - {\e^{2\sigma_0} \over 4\pi G}
\left(6A - C - {4C \over r_0^2R_0} \right) \right\}Q \nn
&& + \left\{ {64\pi b' \sigma_0 R_0 \over 3}\left( -A 
+ {2 \over C}A^2 \right) - {\e^{2\sigma_0} \over 4\pi G}(2A-C) \right\}P \nn
&\equiv& N_Q(A) Q + N_P(A) P\ .
\eea
In order that the above two algebraic equations have  
non-trivial solutions for $P$ and $Q$, $A$ should satisfy 
the following equation
\be
\label{egneq}
0=F(A)\equiv M_Q(A)N_P(A)-M_P(A)N_Q(A)\ .
\ee

Before analyzing Eqs.(\ref{aleq2}), (\ref{aleq1}) and (\ref{egneq}), we 
now briefly discuss how (anti-)evaporation is described.
As in (\cite{NOa}), we consider the following function as an eigenfunction 
of $\Delta$ in (\ref{eigenRS}):
\be
\label{cshS}
f_A(t,r)=\cosh t\alpha\sqrt{C}\cosh^\alpha r\sqrt{C} \ ,\ \ \ 
A\equiv {\alpha (\alpha - 1) C \over 4}\ .
\ee
Note that there is one to one correspondence between $A$ and $\alpha$ if 
we restrict $A>0$ and $\alpha<0$. 
Any linear combination of two solutions is a solution. The 
perturbative equations of motion (\ref{perte}) are always linear 
differential equations. The horizon is given by the condition
\be
\label{horizon}
\nabla\sigma\cdot\nabla\sigma=0\ .
\ee
Substituting (\ref{cshS}) into (\ref{horizon}), 
we find the horizon is given by $r=\alpha t$.
Therefore on the horizon, we obtain 
$S(t,r(t))=Q\cosh^{1+\alpha} t\alpha\sqrt{C}$. 
This tells that the system is unstable if there is a solution $0>\alpha >-1$, 
i.e., $0<A<{C \over 2}$. On the other hand, the perturbation becomes stable 
if there is a solution where $\alpha<-1$, i.e., $A>{C \over 2}$. The 
radius of the horizon $r_h$ is given by 
$r_h=\e^\sigma =\e^{\sigma_0 + \epsilon S(t,r(t))}$.
Let the initial perturbation is negative $Q<0$. Then the radius shrinks 
monotonically, i.e., the black hole evaporates in case of $0>\alpha>-1$.
On the other hand, the radius increases in time and approaches to the 
Nariai limit asymptotically 
$S(t,r(t))\rightarrow Q\e^{(1+\alpha) t |\alpha|\sqrt{C}}$ 
in case of $\alpha<-1$. The latter case corresponds to the anti-evaporation
of black hole observed by Bousso and Hawking \cite{BH}.

We should be more careful in the case of $A={C \over 2}$. When 
$A={C \over 2}$, $f_A(r,t)$ are, in general, given by
\be
\label{solS}
f_A(r,t)= \left\{{\cosh\left((t+a)\sqrt{C}\right) \over 
\cosh \left(r\sqrt{C}\right)}
+\sinh \left(b\sqrt{C}\right)  \tanh \left(r\sqrt{C}\right)\right\}\ .
\ee
Then the condition (\ref{horizon}) gives $t+a=\mp (r-b)$.
Therefore on the horizon, we obtain $S(t,r(t))= Q\cosh b$.
This is a constant, that is, there does not occur evaporation nor 
anti-evaporation. The radius of the horizon does not develop in time.
In classical case, (\ref{egneq}) has the following form
\be
\label{egneqc}
0=\left\{ {1 \over 4\pi G}\e^{2\sigma_0}\left(2A - C \right) \right\}^2 
- {C \over 8\pi G}\e^{2\sigma_0}\times {\e^{2\sigma_0} \over 
4\pi G}
\left(6A - C - {4C \over r_0^2R_0} \right) \ .
\ee
Since $R_0=4r_0^2$ in the classical case, the solution of (\ref{egneqc}) 
is given by $A={C \over 2}$. Therefore the horizon does not develop in
time and the black hole does not evaporate nor anti-evaporate. The result 
is, of course, consistent with that of (\cite{BH}).

In general, it is very difficult to analyze Eqs.(\ref{aleq2}), (\ref{aleq1}) 
and (\ref{egneq}). The equations, however, become simple in the limit of 
$r_0\rightarrow \infty$. Note that $r_0$, the radius of $S_2$ is a free 
parameter and SDW type expansion in Eq.(\ref{SDW}) becomes exact in the 
limit since $R\sim {\cal O}(r_0^{-2})$. In order to consider the limit, 
we now redefine $R_0$ and $\sigma_0$ as follows
\be
\label{reRs}
R_0={1 \over r_0^2}H_0\ \, \ \ \ 
\sigma_0=-\ln (\mu r_0) + s_0\ .
\ee
Then (\ref{aleq2}) is rewritten as follows:
\be
\label{aleq2b}
0=H_0-{4\Lambda \over \mu^2}\e^{2s_0}+4\ .
\ee
Substituting (\ref{aleq2b}) into (\ref{aleq1}), we obtain
\be
\label{aleq1b}
(-H_0^2 +4 )\ln(\mu r_0) + {\cal O}(1)=0\ .
\ee
This tells
\be
\label{H0}
H_0=\pm 2 + {\cal O}\left( (\ln(\mu r_0))^{-1} \right)\ .
\ee
Since we can expect that $H_0\sim 2$ would correspond to the classical 
limit where $H_0=4$, we only consider the case of $H_0\sim 2$. 
Then using (\ref{aleq2b}), we find
\be
\label{s0}
\e^{2s_0}={3\mu^2 \over 2\Lambda}\ .
\ee
Then the metric in the quantum Nariai-type solution has the following form:
\be
\label{qNari}
ds^2={3C \over 2\Lambda}{1 \over \cosh^2\left(r\sqrt{C}\right)}(-dt^2+dr^2)
+{3C \over 2\Lambda}d\Omega
\ee
Substituting (\ref{H0}) and (\ref{s0}) into (\ref{egneq}), we obtain
\bea
\label{FA0}
0&=&F(A) \nn
&=&{1 \over r_0^4}\left\{ \left(\ln (\mu r_0)\right)^2
\left({128\pi b' \over 3}\right)^2\left(-A + {2A^2 \over C}
\right)^2 + {\cal O}(1) \right\} \ .
\eea
Eq.(\ref{FA0}) tells that $A=0, {C \over 2}$. When 
$A=0$, $S$ is constant and there does not occur evaporation nor 
anti-evaporation. As discussed before, the horizon does not develop in time
when $A={C \over 2}$, either.
The result, however, might be an artifact in the limit of $r_0
\rightarrow \infty$. In order to consider physically reliable results, 
we start from the next order of $\left( \ln (\mu r_0) \right)^{-1}$.
Then using (\ref{aleq2}) and (\ref{aleq1}), we find
\bea
\label{Hsnext}
H_0&=&2+\left( \ln (\mu r_0) \right)^{-1}
\left({2b+3b' \over b}+{9 \over 512\pi^2 b G\Lambda}\right) \nn
\sigma_0&=&- \ln (\mu r_0) + {1 \over 2}\ln \left(
{3\mu^2 \over 2\Lambda}\right) \nn && +\left( \ln (\mu r_0) \right)^{-1}
{\mu^2 \over 8\Lambda}\left({2b+3b' \over b}+{9 \over 512\pi^2 b G\Lambda}
\right) \ .
\eea
Since we are now interesting in the problem of anti-evaporation,
we consider the solution of $A\sim {C \over 2}$. The solution  
$A\sim 0$ would correspond to usual evaporation. Assuming
$A={C \over 2}+\left( \ln (\mu r_0) \right)^{-1}a_1$ 
and substituting (\ref{Hsnext}) into (\ref{egneq}), we find
\be
\label{sola1}
a_1=0\ ,\ \ \ a_1=- {(b+b')C \over 8b'}
\ee
In the first solution, the horizon does not develop in time again and we 
would need the analysis of the higher order of $\left( \ln (\mu r_0) 
\right)^{-1}$. An important thing is the second solution is positive 
when $2N+7N_{1/2}>26N_1$. When $a_1$ is positive, $A>{C \over 2}$, i.e., 
there occurs anti-evaporation.
Let us consider $SU(5)$ group with $N_s$ scalar multiplets and 
$N_f$ fermion multiplets in the adjoint representation of gauge group. 
Then, above relation looks as $2N_s+7N_f>26$. 
We see that  for $SU(5)$ GUT with three spinor multiplets and 
 three scalar multiplets 
it is expected anti-evaporation for SdS BH. Similarly one
can estimate the chances for anti-evaporation in the arbitrary 
GUT under discussion.
 On the other hand, when $2N+7N_{1/2}<26N_1$, 
there occurs evaporation.  For example for above GUT with two spinor 
multiplets and two scalar multiplets we expect that matter quantum 
effects induce the evaporation of SdS BH.
This result would be non-perturbative and exact 
in the leading order of $1/N$ expansion. 
Of course, $A$ goes to ${C \over 2}$ in the limit of 
$r_0\rightarrow +\infty$, when the SDW type expansion becomes exact, and 
the black hole does not evaporate nor anti-evaporate in the limit. 
If there is some external 
perturbation, which gives effectively finite $r_0$, however, there may 
occur the anti-evaporation.

\section{Energy Flow and No Boundary Condition}

~~~~~~We now briefly discuss  Hawking radiation. Since the space-time 
which we are now considering is not asymptotically Minkowski, we 
will consider the radial component in the flow 
of the energy $T_{tr}=T_{++}-T_{--}$. 

Usually Einstein equation can be written as
\be
\label{Eineq}
{1 \over 16\pi G}\left( R_{\mu\nu}-{1 \over 2} g_{\mu\nu}R\right)
=T_{\mu\nu}^c + T_{\mu\nu}^q\ .
\ee
Here $T_{\mu\nu}^c$ is the classical 
part of the matter energy momentum tensor, 
which vanishes in the case under consideration , and $T_{\mu\nu}^q$ is 
the quantum part, which we are now interested in. Comparing (\ref{Eineq}) with 
(\ref{cons1}), we find
\be
\label{Tq}
T_{\pm\pm}^q={r_0^2 \over 32\pi G}\e^{2\sigma}
\left\{(\partial_\pm\sigma)^2 - \partial_\pm^2\sigma + 2\partial_\pm\sigma
\partial_\pm\rho\right\}\ .
\ee
Substituting the solution of the perturbation (\ref{cshS}), we find
\be
\label{Tqtr}
T_{tr}^q=-{3\alpha(\alpha+1)\epsilon \over 32\pi G\Lambda}\sinh\left(t\alpha
\sqrt{C}\right)\sinh\left(r\sqrt{C}\right)\cosh^{\alpha-1}\left(r\sqrt{C}
\right) + {\cal O}(\epsilon^2)\ .
\ee
$T_{tr}^q$ is positive when $0>\alpha>-1$ and negative when $\alpha<-1$, 
i.e., there is a flow from the event horizon to the cosmological horizon 
when $0>\alpha>-1$. The direction of the flow is changing when $\alpha<-1$.
They exactly correspond to evaporation ($0>\alpha>-1$) and anti-evaporation
($\alpha<-1$).

Let us turn now to the study of no boundary condition in the evolution of 
black holes. As is known, the cosmological BH does not appear after 
the gravitational collapse of star since the background space-time is not de 
Sitter space but flat Minkowski space. Bousso and Hawking, however, 
conjectured that the cosmological black holes could be pair created by the 
quantum process in an inflationary universe since the universe is similar to 
de Sitter space. They have also shown that no boundary condition 
\cite{HH} determines the fate of the black holes and they should always 
evaporate (for minimal scalars). Now we make a similar analysis for conformal 
matter.  In our case, the analytic continuation to Euclidean space-time is 
given by replacing $t=i\tau$. By the further changing the variables $\tau$ and 
$r$ by 
\be
\label{ucoord}
v=\tau \sqrt{C}\ ,\ \ \ \sin u = {1 \over \cosh r\sqrt{C}}\ ,
\ee
the metric in the quantum Nariai type solution corresponding to (\ref{qNari})
has the following form:
\be
\label{NariaiE}
ds^2={3 \over 2\Lambda}\left(du^2 + \sin^2 u dv^2+d\Omega\right)\ .
\ee
The metric (\ref{NariaiE}) tells that 4 dimensional Euclidean space-time 
can be regarded as the product of two two-sphere $S^2\times S^2$. 
In the Euclidean signature, the operator ${4 \over C}\Delta$ becomes the 
Laplacian on the unit two-sphere $S^2$. 
The nucleation of the black hole is described by cutting two sphere at 
$u={\pi \over 2}$ and joining to it a Lorentzian $1+1$-dimensional 
de Sitter hyperbolid \cite{BH} by analytically continuing $u$ by
$u={\pi \over 2}+i\hat t$ and regarding $\hat t$ as the time coordinate.
In the Euclidean signature, the eigenfunction $f_A(t,r)$ in (\ref{cshS})
is not single-valued unless $\alpha$ is an integer. Therefore $f_A(t,r)$ 
is not adequate when we discuss  the nucleation of the black holes. 
Instead of $f_A(t,r)$, we consider the following eigengfunction 
$\tilde f_A(t,r)$ in the Euclidean signature:
\be
\label{egfn}
\tilde f_A(u,v)=f_0 \cos v \cdot P_\nu^1(\cos u)\ ,\ \ \ 
A={\nu(\nu +1) \over 4}C\ .
\ee
Here $f_0$ is a constant and $P_\nu^1(x)$ is given by the associated 
Legendre function:
\be
\label{aL}
P_\nu^1(x)=\sin\pi\nu (1-x^2)^{1 \over 2}\sum_{n=0}^\infty
{\Gamma (1-\nu +n)\Gamma (2+\nu +n) \over (n+1)!n!}
\left({1-x \over 2} \right)^n\ .
\ee
We can assume $\nu\geq 0$ without any loss of generality.
Note that $\tilde f(u,v)$ vanishes at the south pole $u=0$ since 
$P_\nu^1(x=\cos u=1)=0$. Therefore $\tilde f(u,v)$ is single-valued on 
the hemisphere and does not conflict with the no boundary condition 
\cite{HH}. We should also note that $\tilde f(u,v)$ is not real when 
analytically continuing $u$ by $u={\pi \over 2}+i\hat t$ into Lorentzian
signature but, as in \cite{BH}, we can make $\tilde f(u,v)$ to be real in 
the late Lorentzian time (large $\hat t$) by the suitable choice of 
the constant $f_0$ since
\be
\label{lateP}
P_\nu^1(\cos u)=P_\nu^1(-i\sinh \hat t) \sim {\Gamma\left(\nu + {1 \over 2}
\right) \over \sqrt{\pi} \Gamma (\nu) }(-i)^\nu \e^{\nu \hat t}
\ \ \mbox{when}\ \ \hat t \rightarrow +\infty \ .
\ee
Since $A>{C \over 2}$ when $\nu>1$, we can expect that there would occur 
anti-evaporation in the pair-created black holes. In order to confirm it, 
we consider the behavior of the horizon in the late Lorentzian time (large 
$\hat t$). By using (\ref{lateP}), we find that the condition of 
the horizon (\ref{horizon}) gives $\cos v \propto \e^{-\hat t}$.
Therefore we find on the horizon $\hat f(v(\hat t),\hat t)\propto 
\e^{(\nu -1)\hat t}$.
This tells, as in the case of the previous section, the perturbation is 
stable when $A>{C \over 2}$ ($\nu >1$) and instable when $A<{C \over 2}$ 
($0<\nu <1$). The previous analysis in (\ref{sola1}) implies that there is 
a solution of $A>{C \over 2}$. Therefore the anti-evaporation (stable mode) 
can occur even in the nucleated black holes and some of black holes do not 
evaporate and can survive. This result is different from that of Bousso and 
Hawking, who claimed that the pair created cosmological black holes most 
probably evaporate. (Note however that they considered other type of 
matter-minimal scalars, while we deal with conformal matter). In any case, this question deserves further deep investigation.

\section{Discussion}

~~~~~~In summary, we studied large $N$ effective action for conformal 
matter on spherically symmetric background. Application of this effective 
action to the investigation of quantum evolution of SdS BHs shows 
that such BHs may evaporate or anti-evaporate in nearly degenerated limit.
 No boundary condition is shown to be consistent with  
 anti-evaporation of SdS BHs. Other boundary 
conditions may be discussed in the same way in generalization of this work.
 Some remarks about energy flow 
in regime of evaporation or anti-evaporation are also given.

Let us compare now our results with ones of the previous papers 
\cite{BH,NOa}, where similar questions have been investigated.
In the work by Bousso and Hawking, they treated the quantum effects of 4D 
minimal scalars using $s$-wave and large $N$ approximations. In our previous 
paper \cite{NOa}, the scale ($\sigma$) dependent part of the effective action 
is given by the large $N$ trace anomaly induced effective action (without 
using $s$-wave approximation) but the scale independent part is determined 
using $s$-wave approximation, i.e., spherical reduction. The work 
\cite{NOa} deals with conformal quantum scalars only. In the present 
work, we used the effective action whose scale dependent part is given by 
large $N$ trace anomaly as in the previous paper \cite{NOa} but whose scale 
independent part is given by the Schwinger-De Witt type expansion, which is 
essentialy the power series expansion on the curvature invariants 
corresponding to the rescaled metric (\ref{OVI}). Since the curvature in the 
Nariai limit and the perturbation around it is almost constant, the rescaled 
scalar curvature is always $O(r_0^{-2})$. This tells that the Schwinger-De 
Witt type expansion given in this paper would become exact in the limit of 
$r_0\rightarrow +\infty$. Therefore the analysis given here using 
$r_0\rightarrow +\infty$ limit would be also exact. Moreover, the results 
of present work are given for arbitarary conformal matter (scalars,
spinors and vectors). In all these works, it is
 found the same qualitative result --
the possibility that SdS (Nariai) BH may anti-evaporate. As we see from
 the estimation above such anti-evaporation may be quite general 
for many GUTs. Moreover, pair created (primordial) BHs 
may anti-evaporate due to conformal quantum matter effects when
applying no boundary condition.

As very interesting generalization of above work, it could be helpful 
to understand 
if anti-evaporation is specific feature of SdS BHs or it may be realized 
also for other BHs with multiply horizons. In order to clarify this issue 
we studied Reissner-Nordstr\o m(RN)-de Sitter BHs where preliminary 
investigation shows also the possibility of anti-evaporation due to quantum 
effects. We hope to report on this in near future.

\ 

\noindent
{\bf Acknoweledgments}

We are indebted to A. Sugamoto and 
T. Kadoyoshi for the discussions and the
collaboration at the early stage of this work.



\ 

\noindent
{\Large\bf Appendix}
\appendix

\section{Stringy Presentation of 4D Einstein-scalar Theory}

~~~~~~Starting from Einstein gravity with $N$ minimal scalars,
\be
\label{AI}
S=-{1 \over 16\pi G}\int d^4x \sqrt{-g_{(4)}}
\Bigl[(R^{(4)} -2\Lambda) - {16\pi G \over 2}\sum_{a=1}^N 
g_{(4)}^{\alpha\beta}\partial_\alpha\chi_a
\partial_\beta\chi_a \Biggr]\ .
\ee
one can consider the spherically symmetric space-time:
\be
\label{AII}
ds^2=g_{\mu\nu}dx^\mu dx^\nu +f(\phi)d\Omega\ .
\ee
Reducing the action (\ref{AI}) for the metric (\ref{AII}), we will get
\bea
\label{AIII}
{S \over 4\pi}&=&-{1 \over 16\pi G}\int d^2x\sqrt{-g}\Bigl\{f(R-2\Lambda)
+2+2\left(\nabla^\mu f^{1/2}\right)\left(\nabla_\mu f^{1/2}\right) \nn
&& - {16\pi G \over 2}\sum_{a=1}^N 
f(\phi)\nabla^\alpha\chi_a\nabla_\alpha\chi_a \Biggr\}\ .
\eea
The reduced action belongs to the class of actions described by
\bea
\label{AIV}
S&=&\int d^2x\sqrt{-g}\Bigl\{C(\phi)R+V(\phi)+{1 \over 2}Z(\phi)g^{\mu\nu}
\partial_\mu \phi\partial_\nu \phi \nn
&& - {1 \over 2}\tilde f(\phi)\sum_{a=1}^N 
\nabla^\alpha\chi_a\nabla_\alpha\chi_a \Biggr\}\ .
\eea
where from (\ref{AIII}) we get
\bea
\label{AV}
&& C(\phi)=-{f(\phi) \over 4 G}\ ,\ \ \ 
V(\phi)=-{1 \over 4G}\left(2-2\Lambda f(\phi)\right) \ ,\nn
&&Z(\phi)=-{1 \over 4G}{{f'}^2 \over f}\ ,\ \ \ 
\tilde f(\phi)=-4\pi f(\phi)\ .
\eea
Working in the conformal gauge $g_{\mu\nu}=\e^{2\sigma}\bar g_{\mu\nu}$, 
we may present the action (\ref{AIV}) as a sigma model
\be
\label{AVII}
S=\int d^2x \sqrt{-\bar g}\left[{1 \over 2}G_{ij}(X)\bar g^{\mu\nu}
\partial_\mu X^i\partial_\nu X^j + \bar R\Phi(X) + T(X) \right]
\ee
where
\bea
\label{AVIII}
&& X^i=\{\phi,\sigma,\chi_a\}\ ,\ \ \ \Phi(X)=C(\phi)\ ,\ \ \ 
T(X)=V(\phi)\e^{2\sigma} \ , \nn
&& \nn
&& G_{ij}=\left(\begin{array}{cc|c}
Z(\phi) & 2C'(\phi) & 0 \\ 2C'(\phi) & 0 & 0 \\ \hline 0 & 0 & 
-\tilde f(\phi) \\ \end{array}\right)\ .
\eea
Thus, we presented reduced 4D Einstein-scalar theory as sigma-model. 
Similarly one can present 4D Einstein-conformal scalar reduced theory 
\bea
\label{AIVc}
S&=&\int d^2x\sqrt{-g}\Bigl\{C(\phi)R+V(\phi)+{1 \over 2}Z(\phi)g^{\mu\nu}
\partial_\mu \phi\partial_\nu \phi \nn
&& - {1 \over 2}\tilde f(\phi)\sum_{a=1}^N \left(\nabla^\alpha\chi_a
\nabla_\alpha\chi_a + {1 \over 6}\chi_a^2\right)\Biggr\}\ .
\eea
in a form like (\ref{AVII}) with slightly changed metric $G_{ij}$ 
and $\Phi$, $T$:
\bea
\label{AVIIIc}
&& \Phi(X)=C(\phi)-{1 \over 2}\tilde f(\phi)\ ,\ \ \ 
T(X)=V(\phi)\e^{2\sigma} \ , \nn
&& \nn
&& G_{ij}=\left(\begin{array}{cc|c}
Z(\phi) & 2C'(\phi) & -{1 \over 3}\tilde f'(\phi)\sum_{a=1}^N \chi_a^2 \\ 
2C'(\phi) & 0 & -{2 \over 3}\tilde f(\phi)\chi_a  \\ \hline 
-{1 \over 3}\tilde f'(\phi)\sum_{a=1}^N \chi_a^2 & 
-{2 \over 3}\tilde f(\phi)\chi_a & -\tilde f(\phi) \\ \end{array}\right)\ .
\eea
One can show (see \cite{EO}) that off-shell effective action in stringy 
parametrization (\ref{AVII}) is different from the one calculated in 
dilatonic gravity (\ref{AIV}) in covariant gauge. 
However, on-shell all such
effective actions coincide as it should be (see \cite{EO}). 
The main qualitative result of this Appendix is that 
one can study quantum evolution of black holes using
also sigma-model approach.

\end{document}